\documentclass[11pt,twoside]{article}

\usepackage{asp2006}
\usepackage{epsf}
\usepackage{psfig}
\usepackage{lscape}

\begin{document}
\title{Does the current minimum validate (or invalidate) cycle prediction methods?}
\author{David H. Hathaway}
\affil{NASA Marshall Space Flight Center, Huntsville, AL 35812 USA}

\begin{abstract}
This deep, extended solar minimum and the slow start to Cycle 24 strongly suggest
that Cycle 24 will be a small cycle. A wide array of solar cycle prediction techniques
have been applied to predicting the amplitude of Cycle 24 with widely different results.
Current conditions and new observations indicate that some highly regarded techniques
now appear to have doubtful utility. Geomagnetic precursors have been reliable in the past
and can be tested with 12 cycles of data. Of the three primary geomagnetic precursors only one (the minimum level of geomagnetic activity) suggests a small cycle. The Sun's polar field strength has also been used to successfully predict the last three cycles. The current weak polar fields are indicative of a small cycle. For the first time, dynamo models have been used to predict the size of a solar cycle but with opposite predictions depending on the model and the data assimilation. However, new measurements of the surface meridional flow indicate that the flow was substantially faster on the approach to Cycle 24 minimum than at Cycle 23 minimum. In both dynamo predictions a faster meridional flow should have given a shorter cycle 23 with stronger polar fields. This suggests that these dynamo models are not yet ready for solar cycle prediction.
\end{abstract}

\vspace{-0.5cm}
\section{Introduction}

As each sunspot cycle wanes solar astronomers with widely different interests take their turn at predicting the size and timing of the next cycle. The average length of the previous 22 sunspot cycles is 131.7 months - almost exactly 11 years. However, with one exception, the last 8 cycles have been short cycles with periods closer to 10 years. The minimum preceeding Cycle 23 was in August or September of 1996 so many were expecting the minimum preceeding Cycle 24 to come in 2007 or even 2006. Instead, minimum came in November of 2008 (Fig. 1). This delayed start of Cycle 24, and the depth of the minimum (smoothed sunspot number at its lowest in nearly 100 years) stirred up additional interest and even more predictions \citep{Pesnell08} including talk of an impending grand minimum like the Maunder Minimum \citep[e.g.]{Schatten03}.

\begin{figure}[!ht]
\plotone{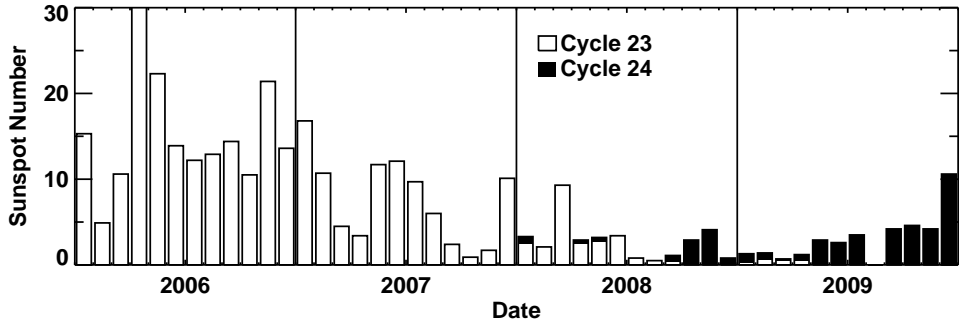}
\caption{
Sunspots associated with Cycle 24 (black) began to dominate over those associated with Cycle 23 (white)in September of 2008. The smoothed sunspot number went through its minimum in December 2008. The smoothed number of spotless days per month went through its maximum in December 2008 as well. The average of these three traditional indicators of sunspot cycle minimum gives November of 2008 as Cycle 24 minimum.}
\end{figure}

\section{Prediction Methods}

Predicting the size and timing of a sunspot cycle is very reliable once a cycle is well underway. Auto-regression techniques \citep{McNish49} and parametric curve fitting techniques \citep{Hathaway94} even give smoothed month-to-month behavior. However, they only become reliable 2-3 years after minimum - at about the inflection point in the rise of the sunspot number toward maximum (Fig. 2).

\begin{figure}[!ht]
\plotone{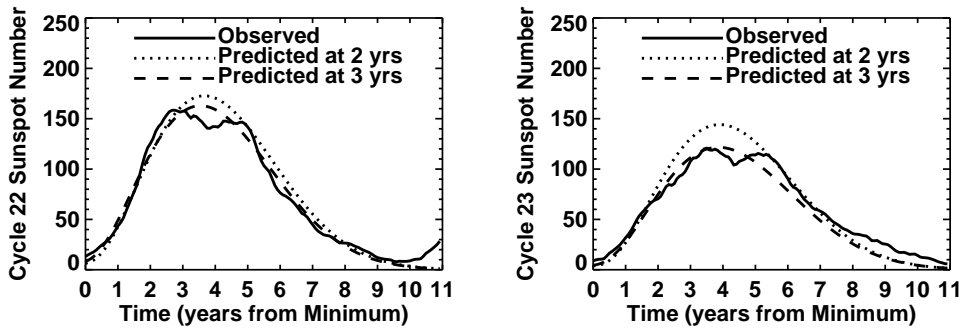}
\caption{
Observed (solid lines) and predicted sunspot numbers for Cycle 22 (left) and Cycle 23 (right). The predictions are based on data taken up to 2 years after minimum (dotted lines) and up to 3 years after minimum (dashed lines) and use the 2-parameter curve-fitting method of \citet{Hathaway94}.}
\end{figure}

Predictions made prior to the start of a cycle or shortly after minimum require methods other than auto-regression or curve-fitting. The simplest method, and the one used as a benchmark for predictive capablilty, is to use an average cycle (maximum smoothed sunspot number $114\pm40$ for cycles 1-23). Many predictions are based on trends or periodicities percieved in the history of cycle amplitudes (eg. the \citet{Gleissberg39} 8-cycle periodicity) . Others are based on the characteristics of the previous cycle or of the cycle minimum itself. In the latter category two characteristics stand out - the Amplitude-Period relation \citep{Wilson98} and the Maximum-Minimum relation \citep{Brown76}.

With the Amplitude-Period relation the amplitude of a cycle is related to the period (length) of the previous cycle - small cycles start late and leave behind a long period cycle. With the Maximum-Minimum relation the amplitude (maximum) of a cycle is related to the level of the minimum preceeding it - small cycles start late and leave behind a low minimum. These two relations are shown in Fig. 3 along with the associated predictions for Cycle 24.

\begin{figure}[!ht]
\plotone{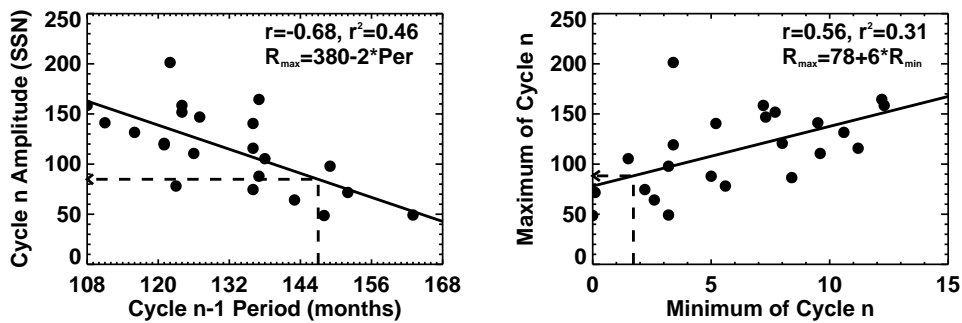}
\caption{
The Amplitude-Period relation (on the left) indicates that big cycles follow short cycles and small cycles follow long cycles (solid line). The 147 month period of Cycle 23 indicates an amplitude of $86\pm30$ for Cycle 24 (dashed lines). The Maximum-Minimum relation (right) indicates that big cycles follow big minima and small cycles follow small minima. The minimum smoothed sunspot number of 1.7 for Cycle 24 minimum indicates an amplitude of $87\pm33$ for Cycle 24.}
\end{figure}

\citet{Hathaway99} examined many of these prediction methods and tested them by backing-up to the minimum predeeding Cycle 19 and using each method to predict Cycles 19-22 but only using data obtained prior to the minimim of each of those cycles. The predictions were examined for both accuracy and stability - stability in the sense of how stable the predicting relation was from cycle-to-cycle. For example, using the average cycle as a predictor gave an RMS error of about 60 and the predicting relation (the size of the average cycle) varied by 12\% from 104 prior to Cycle 19 to 112 prior to Cycle 22. The conclusion from this study was that the most accurate and stable prediction methods were based on geomagnetic precursors - geomagnetic activity near or before the time of sunspot cycle minimum. It was also noted that predictions based on the strength of the Sun's polar fields \citep[see][]{Schatten78} were promising but could not be adequately tested due to the lack of direct data prior to Cycle 21. Since 2006 predictions based on Flux Transport Dynamo Models with assimilated data have been offered - \citet{Dikpati06} and \citet{Choudhuri07}. These three promising methods - Geomagnetic Precursors, Polar Field Precursors, and Flux Transport Dynamos - are examined in the following sections.

\section{Geomagnetic Precursor Predictions}

\citet{Ohl66} was among the first to note that geomagnetic activity around the time of sunspot cycle minimum was a good predictor for the size of the following cycle. In particular, he noted that the minimum in the smoothed monthly geomagnetic index \textit{aa} was well correlated with the amplitude of the following cycle. The \textit{aa} index is a measure of the geomagnetic field variations obtained at 3-hour intervals since 1868 from two nearly antipodal observatories - one in England and one in Austrailia \citep{Mayaud72}. Each observatory was relocated at least once since 1868 and the change made in 1957 seems to have had a significant effect on the data \citep{Svalgaard04}. Fig. 4 illustrates the method described by \citet{Ohl66}. The minima in the smoothed monthly \textit{aa} index are very well correlated with the maximum sunspot number of the following cycle. As of December 2009 the smoothed \textit{aa} index (for June 2009) was still falling and at a record low of 8.8.

\begin{figure}[!ht]
\plotone{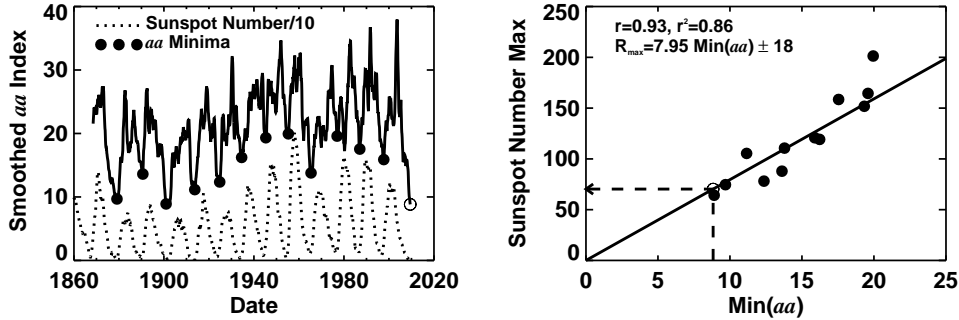}
\caption{
The \citet{Ohl66} geomagnetic precursor method. The minima in the smoothed \textit{aa} index (filled circles on the left) are well correlated (right) with the sunspot number maxima. The current record low (open circles) indicates an amplitude of $70\pm18$ for Cycle 24.}
\end{figure}

It has been known for many decades\citep{Bartels32} that there are two solar sources of geomagnetic activity. One source (now known to be CMEs) has a frequency of occurance that is in phase with the sunspot cycle while the second source (now known to be high-speed solar wind streams) is out of phase with the sunspot cycle and tends to peak late in each cycle. \citet{Feynman82} suggested a method for separating these two components. She noted that background level of geomagnetic activity rose and fell with the sunspot numbers. Removing this sunspot cycle background level of activity leaves behind a component of geomagnetic activity that is out of phase with the sunspot cycle (Fig. 5). \citet{Hathaway99} noted that the peaks in this second component that occur just prior to sunspot cycle minimum were well correlated with the amplitude of the following cycle (Fig. 6). This led \citet{Hathaway06} to a prediction of $160\pm25$ for Cycle 24 based on the assumption that sunspot cycle minimum was eminent in 2006. It is now clear the minimum was still over two years off. The maximum in $aa_I$ used for that prediction was from the fall of 2003 and obviously associated with the 2003 Haloween events. Since this activity was not reflected in significantly higher sunspot numbers, it shows up as a huge (and probably misleading) peak in the $aa_I$ component.

\begin{figure}[!ht]
\plotone{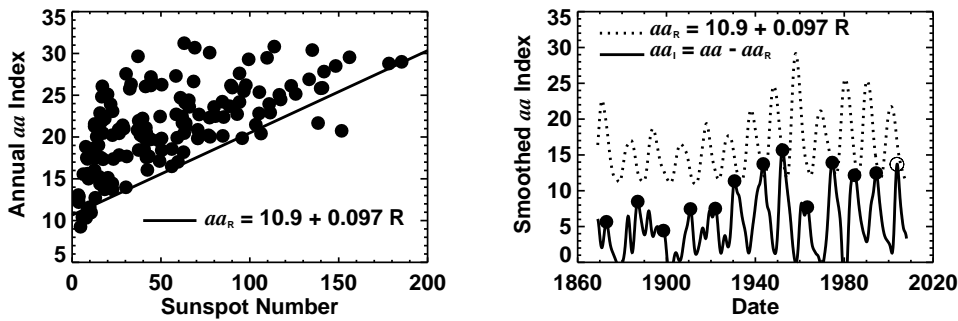}
\caption{
A modified version of the \citet{Feynman82} method for separating geomagnetic activity into a sunspot cycle component $aa_R$ and the remaining ``Interplanetary'' component $aa_I$. A line (the solid line in the left-hand figure) is fit through the lowest points. This determines the sunspot cycle component shown by the dotted curve on the right. The remaining geomagnetic activity is the Interplanetary component shown by the solid curve on the right. The peaks in this component prior to sunspot cycle minimum are shown by the filled circles on the right.}
\end{figure}

\begin{figure}[!ht]
\plotone{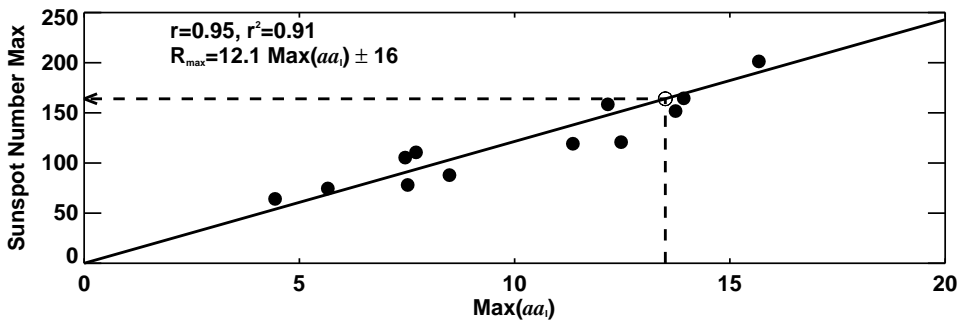}
\caption{
The relationship between the maximum in $aa_I$ just prior to minimum and the following cycle sunspot number maximum. The two quantities are highly correlated. The $aa_I$ maximum of 13.5 found in late 2003 indicates an amplitude of $165\pm16$ for Cycle 24. }
\end{figure}

\begin{figure}[!ht]
\plotone{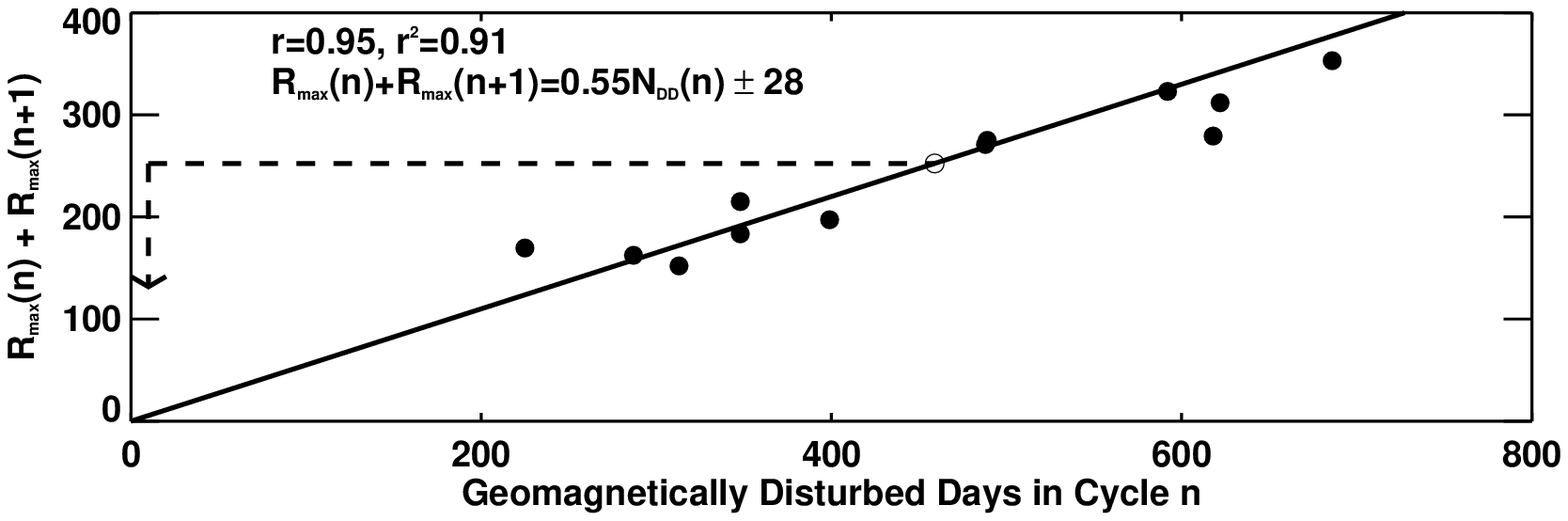}
\caption{
The \citep{Thompson93} relationship between the number of geomagnetically disturbed days in a cycle and the sum of the amplitudes of the current and future cycle. The relationship is highly significant and indicates an amplitude of $130\pm28$ for Cycle 24.}
\end{figure}

\citet{Thompson93} also recognized that some geomagnetic activity in a sunspot cycle was indicative of the amplitude of the following cycle. However, instead of trying to separate the geomagnetic activity into components he found that the total number of geomagnetically disturbed days (defined as days with geomagnetic index $Ap \geq 25$) during a cycle was proportional to the sum of the amplitudes of the current cycle and the future cycle. This is shown in Fig. 7. Here again, as with the Feynman Method prediction, the Halloween events of 2003 significantly impact the process and results. Removing these events lowers the predicted amplitude of Cycle 24 from $130\pm28$ to $95\pm28$.

When we extend the prediction method testing of \citet{Hathaway99} to include Cycle 23 we still find that these Geomagnetic Precursor methods are substantially better than other methods. Although this testing indicates that the Thompson and Feynman Methods faired slightly better than the Ohl Method for Cycles 19-23, the impact of the Halloween 2003 events on those methods suggest that greater weight should be given to the Ohl Method prediction for Cycle 24 - an amplitude of $70\pm18$.

\section{Polar Field Precursor Predictions}

The strength of the Sun's polar magnetic fields near sunspot cycle minimum has been used to predict the last three cycles - Cycle 21 \citep{Schatten78}, Cycle 22 \citep{Schatten87}, and cycle 23 \citep{Schatten96}, with considerable success. This method is based on the dynamo model described by \citet{Babcock61} and \citet{Leighton69} in which the Sun's poloidal field at minimum is amplified and converted into the toroidal field (that erupts in active regions) by differential rotation. While several questions remain about the implementation of this method (What precise phase of the solar cycle should the measurement be taken? Is the relationship between polar fields and sunspot cycle amplitude linear?) the success with predicting the last three cycles places this method on par with the geomagnetic precursor methods. The polar fields as measured at the Wilcox Solar Observatory (Fig. 8) have remained substantially weaker since 2004 leading to a prediction of $78\pm8$ for the amplitude of Cycle 24 \citep{Svalgaard04}.

\begin{figure}[!ht]
\plotone{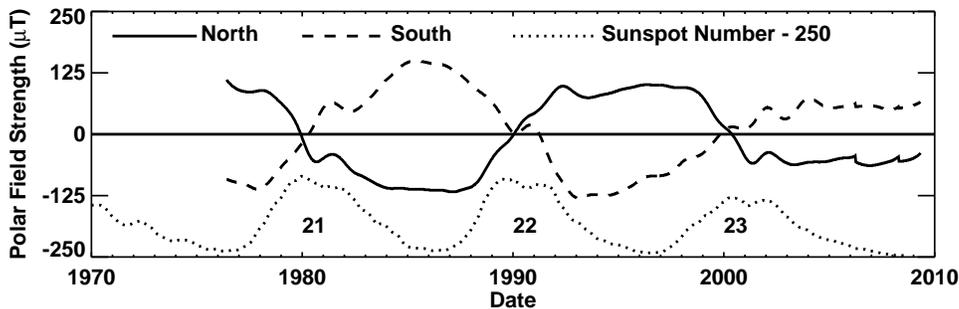}
\caption{
The Sun's polar fields as measured at the Wilcox Solar Observatory. The weakened polar fields seen on the approach to Cycle 24 minimum indicate a weak Cycle 24.}
\end{figure}

\section{Flux Transport Dynamo Predictions}

In a ground-breaking paper \citet{Dikpati06} used a dynamo model with assimilated data to predict a solar cycle. Their dynamo model is a kinematic flux transport dynamo in which the axisymmetric flows in the convection zone (differential rotation and meridional circulation) are prescribed along with a diffusivity (representing the effects of the non-axisymmetric convective flows) and a field regenerating term (representing the stretching and twisting of magnetic field lines by instabilities and the effect of rotation on rising magnetic flux tubes). Historical sunspot cycle data was assimilated into the model by adding magnetic sources at the surface representative of observed sunspot areas and positions. The strength of the toroidal fields in the model were found to accurately reflect the strength of the last 10 sunspot cycles. They concluded that Cycle 24 would be 30-50\% larger than cycle 23 i.e. an amplitude of 160-180 for Cycle 24. They went on to note that the speed of the meridional flow had apparently slowed as Cycle 23 approached maximum \citep{Basu03}. In their model a slow meridional flow produces long cycles and weak polar fields. From this they concluded that Cycle 24 would start late.

Shortly after the publication of this paper \citet{Choudhuri07} presented their own prediction based on a similar Flux Transport Dynamo model. Their model had one substantial difference from the \citet{Dikpati06} model - a significantly larger diffusivity. In addition, instead of assimilating sunspot area data they reset the poloidal field at minimum for the last three cycles and found a good fit to the observed cycle amplitudes. Putting in the weak polar fields at the current minimum predicted a Cycle 24 about 35\% weaker than Cycle 23 - an amplitude of about 80 for Cycle 24 - right in line with the Polar Field Precursor prediction of \citet{Svalgaard04}. 

\section{Meridional Flow Variations}

In a recent paper \citet{Hathaway10} measured the changes in the speed of the surface meridional flow over the completed Cycle 23. Over 60,000 fulldisk magnetograms from the MDI instrument on SOHO were used to determine the meridional motion of weak magnetic features that are carried by the flow. They found that while the speed of the meridional flow did indeed slow on the approach to Cycle 23 maximum in 2000/2001, it then sped up to  substantially faster speeds for the remainder of the cycle (Fig. 9). This type of variation was also seen in Cycles 21 and 22 by \citet{Komm93}. However the faster speed on the approach to cycle 24 minimum should have produced \textit{stronger} polar fields and a \textit{shorter} cycle 23 with the Flux Transport Dynamo models.

\begin{figure}[!ht]
\plotone{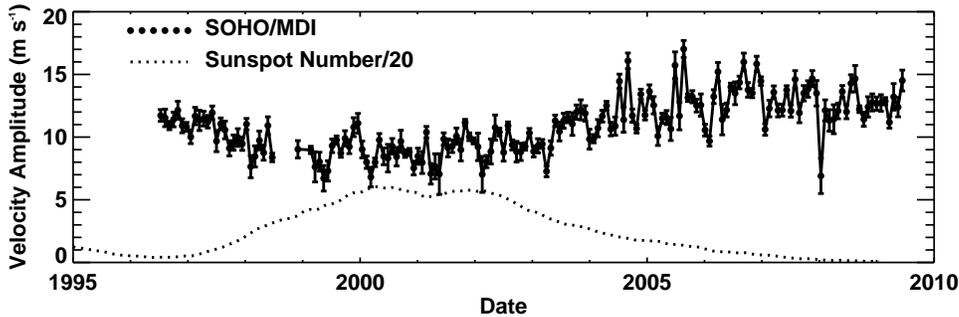}
\caption{
The meridional flow speed averaged over individual solar rotations during Cycle 23 (dots with 2-sigma error bars). The flow slows on the approach to maximum but then speeds up to substantially higher speeds.}
\end{figure}
\section{Conclusions}

Ohl's Geomagnetic Precursor, the Polar Field Precursors, the Amplitude-Period relation, and the Maximum-Minimum relation all indicate that Cycle 24 will be small with an amplitude of about 75. The other two geomagnetic precursor methods appear to be unduly impacted by the activity associated with the Halloween events of 2003 and give larger cycles. We conclude with \citet{Wang09} that the more appropriate geomagnetic precursor is that of \citet{Ohl66} - the minimum level of geomagnetic activity.

The predictions based on Flux Transport Dynamos gave very different predictions but they both predict behavior in conflict with the observed meridional flow variations. The faster meridional flow after Cycle 23 maximum sould give a short cycle with strong polar fields acording to these models. Instead we find a long cycle with weak polar fields. We must conclude with \citet{Tobias06} that these dynamo modes are not yet ready for cycle predictions.

\end{document}